\documentclass[12pt]{article}
\usepackage{a4,amsmath,epsfig,amssymb}
\usepackage[latin1]{inputenc}
\usepackage{graphicx}
\usepackage{caption}
\usepackage{subcaption}
\usepackage{feynmp}
\usepackage{amsmath}
\begin{document}
\bigskip\bigskip

\begin{center}
\begin{large}
{\large \bf Parametrizations of the Spin Density Matrix}
\end{large}
\end{center}

\vspace{8pt}
\begin{center}
\begin{large}
  
E.~Di Salvo$^{a,b,}$\footnote{Elvio.Disalvo@ge.infn.it}, Z.~J.~Ajaltouni$^{a,}$\footnote{ziad@clermont.in2p3.fr}
\end{large} 

\bigskip

$^a$ 
Laboratoire de Physique Corpusculaire de Clermont-Ferrand, \\
IN2P3/CNRS Universit\'e Blaise Pascal, 
F-63177 Aubi\`ere Cedex, France\\

\noindent  
$^b$ 
Dipartimento di Fisica dell' Universita' di Genova \\
and I.N.F.N. - Sez. Genova,\\
Via Dodecaneso, 33, 16146 Genova, Italy \\  

\noindent  

\vskip 1 true cm
\end{center}
\vspace{1.0cm}

\vspace{6pt}
\begin{center}{\large \bf Abstract}

We propose for the spin density matrix two parametrizations which automatically fulfil the non-negativity conditions, without setting any bound on the parameters. The first one relies on a theorem, that we prove, and it is rather simple and easily adaptable to some specific reactions, where, for example, parity is conserved or angular momentum conservation entails selection rules. Moreover, in the case when the rank is less than the order of the density matrix, we show how to improve the fits to the data, either by implementing previous suggestions, or by elaborating an alternative method, for which we prove a second theorem. Our second parametrization is a variant of previous treatments, it appears suitable for some particular processes. Moreover, we discuss about the possibility of inferring the elements of the density matrix from the differential decay width. Last, we illustrate various examples of current interest, both in strong and weak decays; some of them may be helpful in the investigation of physics beyond the standard model.
\end{center}

\vspace{10pt}

\centerline{PACS numbers: PACS Nos.: 23.20.En, 23.20.Js, 23.40.Bw, 24.10.-i}

\newpage

\section{Introduction}

The spin density matrix (SDM), which was introduced long time ago\cite{fa} to describe a mixture of pure spin states, is an essential tool for various aims, like determining the spin and the parity of the resonances[2-9], singling out some exchange mechanisms in the production reactions[8,10-19] and finding possible hints to new physics[20-22]. It concerns unstable states - to be denoted as $R$ in the following - which are produced in some reactions, usually of the type 
\begin{equation} 
a ~~~ b ~~\to~~ R ~~~ c ~~~~ \mathrm{or} ~~~~ R_0 ~~\to~~ R ~~~ d. \label{rcts}
\end{equation} 
The matrix elements of the SDM of $R$ are inferred through the analysis of one of its decay modes.	

This matrix is Hermitian, non-negative definite and has a unitary trace. It is characterized by a set of $N$ pure, orthonormal spin states $|n\rangle$ - its eigenvectors - such that\cite{fa,min}  
\begin{equation}
\rho = \sum_{n=1}^N |n\rangle p_n\langle n|, 
\ ~~~~~~ \ 0\leq  p_n \leq 1, \ ~~~~~~ \ \sum_{n=1}^N p_n =1.
\label{pos}
\end{equation}
Here the $p_n$, some of which may be zero\cite{eg}, are the probabilities of the various eigenstates. If $R$ consists of a single resonance with spin $J$, one has $N$ = $2J+1$. However, it may happen to cope with an intermediate state that consists of two spins[23-25], of the $\rho$- and $\omega$-resonances\cite{or1,or2} or of a resonance and the background\cite{cht}; in this case, it results $N$ = $\sum_k (2J_k+1)$, $J_k$ being the value of each spin. 

The number of non-zero eigenvalues is defined as the {\it rank} of the matrix, denoted as $r$, with $r$ $\leq$ $N$. If the mechanism, which gives rise to $R$, is of the type (\ref{rcts}), an upper bound to $r$ can be fixed\cite{eg,bls}: obviously, this bound is important only if it is less than the order of the SDM. 

In a frame at rest with respect to $R$, the eigenstates $|n\rangle$ are not necessarily eigenstates of the operators ${\bf J}^2$ and $J_z$; as an example, for a spin-3/2 resonance, the SDM may be diagonal with respect to the states 
\begin{equation} 
|3/2, 3/2\rangle, ~ |3/2, 1/2\rangle, ~  a|3/2, -1/2\rangle + 
b|3/2, -3/2\rangle, ~ \mathrm{and} ~ b|3/2, -1/2\rangle 
- a|3/2, -3/2\rangle,
\end{equation}
with $|a|^2+|b|^2$ = 1.
If $R$ involves a single spin, a given eigenstate of $\rho$ can be reduced to an eigenstate of $J_z$ by a rotation of the reference frame, but the unitary matrix that diagonalizes the SDM is not a rotation in the general case\cite{ld}. 

The SDM is generally characterized by $N^2-1$ real parameters. However, if parity is conserved in the production mechanism\cite{da,cht}, this number is reduced, owing to some relations between the matrix elements; moreover, if parity is conserved also in the successive decay of $R$, it often happens that the imaginary part of $\rho$ cannot be measured\cite{da,cht,dd}; last, if $R$ is produced in a decay, the constraint of angular momentum conservation\cite{ab1,ab2,aj} has to be accounted for.

Obviously, all constraints that we have exposed above, including non-negativity\cite{min,dz,da,cht,bls,dd} and normalization condition, have to be fulfilled by a parametrization of the SDM. A problem, which experimental physicists are coped with, is to avoid introducing any external bounds on the parameters. Different solutions were proposed in past years\cite{zw,da,cht}; however, a simple and  sufficiently flexible parametrization seems to be still lacking\cite{dd}. Moreover, the suggestion of exploiting the rank condition for reducing the number of parameters of the SDM\cite{min,cht} needs implementation; this problem is especially important when higher spins or more than one spin are involved. Last, the parametrizations, which were proposed in the past, concern essentially resonances which are produced in strong interactions, whereas, in the last years, one has to do with structures which arise from weak decays\cite{ab1,ab2,dsa}: a systematic study of such cases has not yet been performed.  

The aim of the present paper is to fill such gaps. In particular, we consider the SDM both for parity conserving and for parity violating processes; furthermore, we suggest two different parametrizations that automatically satisfy the above mentioned constraints. The first parametrization, which is based on a theorem, is simple and easily adaptable to different situations; moreover we implement and suggest some methods for recognizing possible null eigenstates and for reducing the number of independent parameters. The second parametrization is inspired by an unusual method\cite{bls,chp,aml,per,zh}, which takes into account the mechanism that originates the structures we want to study: indeed, in some cases, it may be convenient to parametrize the SDM as\cite{bls} 
\begin{equation}
\rho = U\rho^{(i)}U^{\dagger},  \label{evol}
\end{equation}
where $\rho^{(i)}$ is the initial density matrix and $U$ a unitary operator that describes the evolution of the reaction. 

Moreover, we discuss about the possibility of obtaining the elements of the SDM from data, referring in particular to  the case when more spins are involved, or when parity is conserved in the production mechanism. Last, we apply our parametrizations to some reactions of interest.

Sect. 2 is dedicated to the first parametrization, for which we prove a preliminary theorem. In Sect. 3, we show how to exploit situations such that $r$ $<$ $N$, in part by using a second theorem. In Sect. 4, we introduce the second parametrization. Sect. 5 is devoted to a discussion about the extraction of the SDM elements from the differential decay width. Last, we illustrate a few examples in Sect. 6 and draw some conclusions in Sect. 7.  

\section{Parametrization of the Spin Density Matrix - I}

The first parametrization is based on a theorem, which we state and prove preliminarily.

\subsection{Theorem}

''Consider an $N\times N$ Hermitian matrix, $\rho$, defined with respect to an orthonormal basis,
\begin{equation}
\rho = \sum_{i,j=1}^N |i\rangle \rho_{ij}\langle j|. \label{rho}
\end{equation}
A necessary and sufficient condition for it to be non-negative is that

- all of its diagonal elements are non-negative and

- the Schwarz inequality\cite{da} 
\begin{equation}
|\rho_{ij}|^2\leq \rho_{ii}\rho_{jj} \label{sch}
\end{equation}
holds.''

{\it Proof} 
  
a) The necessary condition is a consequence of the features of the characteristic equation of a non-negative definite Hermitian matrix\cite{cht}. However, we give a different argument, which is similar to
the one proposed by Daboul\cite{da}.
If $\rho$ is Hermitian and non-negative, we may set 
\begin{equation} 
\rho_{ij} = \sum_n U^*_{in} U_{jn} p_n, ~~~~ \sum_n U^*_{in} U_{jn} = \delta_{ij}. \label{sdm}
\end{equation}
Here $U$ is a unitary matrix\footnote{As observed by Daboul\cite{da}, such a matrix is defined up to some given phases, $U_{jn}$ $\to$ $U_{jn} e^{i\zeta_n}$} that diagonalizes  $\rho$ and $p_n \geq 0$ is the $n$-th eigenvalue of $\rho$. Then 
\begin{equation} 
\rho_{ii} = \sum_n |U_{in}|^2 p_n ~~ \geq 0, \label{pod}
\end{equation}
which proves the non-negativity of the diagonal elements of $\rho$. Moreover, Eq. (\ref{sdm}) suggests to define a set of complex vectors\cite{da}: 
\begin{equation} 
|V_i\rangle = \sum_n U_{in} \sqrt{p_n}|n\rangle. \label{cvct}
\end{equation}
Then, owing to the first Eq. (\ref{sdm}), one has $\rho_{ij}$ = $\langle V_i|V_j\rangle$ and the Schwarz inequality for the scalar product implies (\ref{sch}).

b) Now suppose that $\rho$ is Hermitian, with non-negative diagonal elements and satisfies the condition (\ref{sch}). Therefore, we may regard each matrix element $\rho_{ij}$ as a scalar product:
\begin{equation} 
\rho_{ij} = \langle W_{i}|W_{j}\rangle, \label{scpr1}
\end{equation}
where $|W_i\rangle$ ($i$ = 1, 2, ... $N$) is a set of complex vectors. This set may be fixed in infinitely many ways. Indeed, we may decompose each vector $|W_i\rangle$ according to the orthonormal basis adopted for defining the elements of the SDM, Eq. (\ref{rho}), {\it i. e.},

\begin{equation} 
|W_i\rangle = \sum_{k=1}^N \alpha^k_i |k\rangle.   
\end{equation}
Then, Eq. (\ref{scpr1}) entails the system 
\begin{equation} 
\rho_{ij} = \sum_{k=1}^N {\alpha^k_i}^* \alpha^k_j, 
\end{equation} 
which is undetermined, as it consists of $N^2$ real equations and $2N^2$ real unknowns. We give in Appendix A one of the infinite solutions to the system. 

Given any vector
\begin{equation} 
|Y\rangle = \sum_{i=1}^N y_i |i\rangle,
\end{equation}
Eq. (\ref{scpr1}) implies
\begin{equation} 
\langle Y|\rho|Y\rangle = \sum_{i,j=1}^N y^*_i\langle W_i|W_j \rangle y_j =\langle Z|Z \rangle \geq 0,
\end{equation}
where
\begin{equation} 
|Z \rangle = \sum_{i=1}^N y_i |W_i\rangle. 
\end{equation}
This shows the non-negativity of $\rho$ and completes the proof of our theorem.

\subsection{Parametrization}

The results of the previous subsection are now exploited for parametrizing the SDM of the unstable state $R$, which, as already explained, may consist of more spins.

Since the diagonal elements of the SDM are non-negative, we introduce $N$ real parameters $a_i$, such that 
\begin{equation} 
\rho_{ii} = a_i^2 ~~~~ \mathrm{and} ~~~~ \sum_{i=1}^N a_i^2 = \sum_{i=1}^N \langle W_{i}|W_{i}\rangle = 1. \label{diag}
\end{equation}
Moreover, we observe, analogously to Doncel {\it et al.}\cite{dmm3}, that the set of the moduli of the $N$ vectors (\ref{cvct}) - which characterize the SDM - can be related to the points of the surface of a hypersphere of unitary radius in the $N$-dimensional Euclidean space $R^N$. Therefore, we introduce generalized spherical coordinates in that space, by means of a number of 'angular' parameters $\alpha_i$. We propose a parametrization for the SDM of $R$, distinguishing between parity violation and parity conservation in the processes (\ref{rcts}).

\subsubsection{Parity Violating Processes}

We define, in this case, $N-1$ 'angular' parameters:
\begin{eqnarray} 
a_1 &=& cos\alpha_1, ~~~~~~ a_2 = cos\alpha_2 sin\alpha_1, ~~~~ ... \\
a_i &=& cos\alpha_i \Pi_{l=1}^{i-1} sin\alpha_l ~~~ ... ~~~ a_N = \Pi_{l=1}^{N-1} sin\alpha_l. \label{parsdm}
\end{eqnarray}   

As regards the off-diagonal matrix elements, the Schwarz inequality (\ref{sch}) suggests to set
\begin{equation} 
\rho_{ij} = |a_i| |a_j| cos\gamma_{ij} e^{i\phi_{ij}}. \label{psch}
\end{equation}

\subsubsection{Parity Conserving Processes}

In this case, we fix a plane, say $\pi$, to be identified with the production plane\cite{cht} or, in the case of the decay, with the plane which is singled out by the momenta of $R_0$ and $R$. Two possible choices are available\cite{chu}, corresponding to fixing the quantization axis respectively normal to $\pi$ or lying on it. Here we focus on the latter case, which includes the helicity representation. We shall see two examples in sect. 6.
The matrix elements of the SDM are denoted as $\rho^{JJ'}_{mm'}$. Then, fixing the $y$-axis normally to the $\pi$-plane, and defining the reflection operator $\Pi_y$ = $P exp(-i\pi J_y)$, where $P$ is the parity operator, one has 
\begin{equation} 
\Pi_y |J m\rangle = \eta e^{-i\pi(J-m)} |J -m\rangle; \label{stpp}
\end{equation}
here $\eta$ is the intrinsic parity of the state that we consider. Therefore
\begin{equation} 
(\Pi_y \rho \Pi_y^{-1})^{JJ'}_{mm'}= \eta\eta' e^{-i\pi\Delta}\rho^{~J~~J'}_{-m-m'},  ~~ \mathrm{with} ~~  
\Delta = J-J'-m+m', \label{ppc}
\end{equation} 
$J$ and $J'$ being different spin values and $m$ and $ m'$ their third components. Parity invariance implies	
\begin{equation} 
\rho^{JJ'}_{mm'} = \eta\eta' e^{-i\pi\Delta}\rho^{~J~~J'}_{-m-m'}. \label{yr}
\end{equation}

As regards the parametrization of the diagonal elements of the SDM, we distinguish two cases:

a) Odd $N$, corresponding to an odd number of integer spins:
\begin{eqnarray} 
a_1 &=& a_N = \frac{1}{\sqrt{2}}cos\alpha_1, ~~~~~~ a_2 = a_{N-1} = \frac{1}{\sqrt{2}}cos\alpha_2 sin\alpha_1, ~~~~ ...\\
  a_i &=& = a_{N-i+1} = \frac{1}{\sqrt{2}} cos\alpha_i \Pi_{l=1}^{i-1} sin\alpha_l ~~~ ... ~~~ a_{N'/2}= \Pi_{l=1}^{N'/2} sin\alpha_l, \label{parcv1}
\end{eqnarray}
with $N'$ = $N-1$.

b) Even $N$, in all other cases:
\begin{eqnarray} 
a_1 &=& a_N = \frac{1}{\sqrt{2}}cos\alpha_1, ~~~~~~ \ ~~~~~~ a_2 = a_{N-1} = \frac{1}{\sqrt{2}}sin\alpha_1 cos\alpha_2, ~~~~ ...\\
  a_i &=& = a_{N-i+1} = \frac{1}{\sqrt{2}} cos\alpha_i \Pi_{l=1}^{i-1} sin\alpha_l ~~~ ... ~~~ a_{N/2}= a_{N/2+1} = \frac{1}{\sqrt{2}}\Pi_{l=1}^{N/2} sin\alpha_l. \label{parcv2}
\end{eqnarray}

The off-diagonal elements are parametrized according to Eq. (\ref{psch}), taking account of the relationship (\ref{yr}).  

If the imaginary part of $\rho_{ij}$ is not measurable, as we shall discuss in Sect. 5, one has to set $\phi_{ij}$ = 0 in the parametrization (\ref{psch}); incidentally, we note that the condition $(\Re\rho_{ij})^2$ $\leq$ $|\rho_{ii}\rho_{jj}|$, and therefore the Schwarz inequality - which is necessary to guarantee the non-negativity condition - is automatically fulfilled; moreover, $|\Re\rho_{ij}|$ constitutes a natural lower bound to $|\rho_{ij}|$, without introducing any external constraint\cite{da}. One can also apply the rank condition, according to which
\begin{equation} 
r(\Im \rho) \leq r(\Re \rho) - r(\rho),
\end{equation}
where $\Re \rho$ and $\Im \rho$ are respectively the real and imaginary part of $\rho$\cite{cht}.

\section{Exploiting the Rank of the SDM}

The complete parametrization of the SDM requires, in principle, all of the parameters that we have illustrated in the previous section. However, in some specific situations, considerable simplifications are possible, which are notably useful in the cases of high spins or of more states. Here we examine the case when the rank of the SDM is less than its order, that is, when some of the eigenvalues $p_n$ - or equivalently the determinant and some of the principal minors - vanish.  

Therefore, an important problem is to single out the kernel of the SDM\cite{eg,pe,min,cht}. To accomplish that, one has to proceed according to more steps\cite{cht}. At first, one fits the experimental data by means of an order-$N$ SDM, then one determines its eigenvalues, $p_n$, and its eigenvectors, $|n\rangle$. If some of the $p_n$ vanish, the $N$-dimensional space where the SDM acts may be divided into two subspaces, the kernel, of dimension $N_k < N$, and the complementary one, whose basis is constituted by the remaining $r$ = $N-N_k$ eigenvectors. As regards the successive steps, we indicate three different methods.

{\it Method 1}

Let     
\begin{equation} 
|k\rangle = \sum_{i=1}^N \alpha_{ki} |i\rangle, ~~ k = 1,2, ... N_k \label{kre}
\end{equation}
be the eigenvectors of the kernel: they are expressed with respect to the basis that has been adopted in Eq. (\ref{rho}). The matrix
\begin{equation} 
\rho' = U\rho U^{\dagger}, \label{tdm}
\end{equation}
with
\begin{eqnarray} 
U_{ki} &=& \delta_{ki} ~~ \mathrm{for} ~~ 1\leq k\leq r, \\
 &=&\alpha^*_{ki} ~~ \mathrm{for} ~~ r+1\leq k\leq N, 
\end{eqnarray}
results in
\begin{equation} 
\rho'_{kl} = \rho_{kl} = \eta_{kl} ~~ \mathrm{for} ~~ 1\leq ~ k,l ~ \leq r, ~~ 0 ~~ \mathrm{otherwise}. \label{subm}
\end{equation}
Moreover, the $r\times r$ sub-matrix $\eta$, which is defined by Eq. (\ref{subm}), is non-singular; it can be re-parametrized by using the procedure that we have described in the previous section. 

{\it Method 2}

If the unstable state $R$ has a fixed spin $J$, a single eigenstate of $\rho$ can be reduced to an eigenstate of $J_z$ by means of a rotation of the reference frame\cite{ld}. Indeed, any normalized vector, say, 
\begin{equation} 
|l\rangle = \sum_m\alpha^J_m|J,m\rangle, ~~~~ \mathrm{such ~~ that} ~~~~ 
\sum_m |\alpha^J_m|^2 = 1,
\end{equation}
may be re-written as 
\begin{equation}
|l\rangle = U[R(\phi,\theta,0)]|J,\bar{m}\rangle = e^{-i\phi J_z} 
e^{-i\theta J_y}|J,\bar{m}\rangle,
\end{equation}
where $|J,\bar{m}\rangle$ is a pure spin state and $\phi$ and $\theta$ suitable values of the azimuthal and polar angle respectively. If $|l\rangle$ is an eigenvector that corresponds to a null eigenvalue, the transformed SDM has at least a
vanishing row and column, which intersect in the main diagonal.

In this connection, we remark that, if the polarization direction of the resonance may be determined - {\it e. g.}, through an asymmetry in a weak decay mode -, it is convenient to rotate the reference frame so as to take the quantization axis along such a direction: the transformed SDM is diagonal.  

{\it Method 3}

The third method for reducing the order of the SDM is based on a simple theorem.

{\it Theorem}: ''If the Schwarz inequality is saturated for some ($i$,$j$)-pair of indices, $i$ $\neq$ $j$, {\it i. e.},
\begin{equation}
|\rho_{ij}|^2 = \rho_{ii}\rho_{jj},\label{sscw}
\end{equation}
the rank of the SDM is at least one unit less than its order.''

{\it Proof} 

We define the $2\times2$ matrix $\bar{\rho}$ as
\begin{equation}
\bar{\rho}_{11} = \rho_{ii}, ~~~~ \bar{\rho}_{22} = \rho_{jj}, ~~~~ \bar{\rho}_{12} = \bar{\rho}^{~*}_{21} = \rho_{ij}.
\end{equation}
The eigenvalues of $\bar{\rho}$ are 0 and $\rho_{ii}+\rho_{jj}$; the corresponding eigenvectors are
\begin{equation}
|n_1\rangle = {\cal N}(\sqrt{\rho_{jj}}|i\rangle -\sqrt{\rho_{ii}}e^{-i\phi_{ij}}|j\rangle) ~~~~ \mathrm{and} ~~~~ 
|n_2\rangle = {\cal N}(\sqrt{\rho_{ii}}|i\rangle +\sqrt{\rho_{jj}}e^{-i\phi_{ij}}|j\rangle),
\end{equation}
where $\phi_{ij}$ is the phase of $\rho_{ij}$ and ${\cal N}=(\rho_{ii}+\rho_{jj})^{-1/2}$. The matrix $\bar{\rho}$ is diagonalized by the unitary matrix ${\cal U}$ such that
\begin{equation}
{\cal U}_{11} = {\cal N}\sqrt{\rho_{jj}}, ~~ {\cal U}_{12}=-{\cal N}\sqrt{\rho_{ii}}e^{-i\phi_{ij}}, ~~ {\cal U}_{22}= {\cal N}\sqrt{\rho_{ii}},
~~ {\cal U}_{21}= {\cal N}\sqrt{\rho_{jj}}e^{-i\phi_{ij}}.
\end{equation}
Therefore, defining the $N\times N$ unitary matrix $U$ as 
\begin{equation}
U_{ii}={\cal U}_{11}, ~~ U_{ij}={\cal U}_{12}, ~~ U_{ji}={\cal U}_{21}, ~~ U_{jj}={\cal U}_{22}, ~~ U_{lm}= \delta_{lm} ~~ \mathrm{for} ~~ l,m \neq i,j,
\end{equation}
it follows that the new SDM $\rho'$ = $U\rho U^{\dagger}$ has at least a diagonal term which vanishes. But as shown before this implies that $\rho'_{ij}$ = $\rho'_{ji}$ = 0 for a fixed $i$ and all $j$ and $r(\rho)\leq N-1$. This completes the proof and indicates how to construct, in this case, the unitary matrix for obtaining the reduction of the parameters.

This result could be generalized: any null principal minor\cite{cht} of the SDM corresponds to at least one zero eigenvalue. In particular, the most trivial principal minor, a zero in the main diagonal, implies the vanishing of the corresponding row and column, as follows automatically from Eqs. (\ref{diag}) and (\ref{psch}).

As a conclusion of this section, we remark that, when the rank of the density matrix is less than its order, the Eberhard-Good theorem\cite{eg,pe,min} can be applied.  

\section{Parametrization of the Spin Density Matrix - II}

The procedure, that we have just described, is especially useful in cases when it is  difficult to determine {\it a priori} an  upper bound to the rank (for example, in inclusive reactions\cite{ad1,ad2}), or when this bound is greater than the order of the SDM\cite{bls}. If the bound is less than the order, or if the spin structures of the initial and final state are particularly simple\cite{te}, it may be sometimes convenient to use an alternative parametrization, based on Eq. (\ref{evol})\cite{per,bls,te}. Indeed, in a process of the type (\ref{rcts}), this equation yields for the SDM of $R$
\begin{equation}
\rho_{kk'} = \sum_m\sum_{ll'} U^m_{kl}\rho^{(i)}_{ll'} U^{m*}_{k'l'}. \label{rhoR}
\end{equation}
Here	
\begin{equation}
U^m_{kl} = \frac{1}{{\cal N}_u}A^m_{kl},  ~~~~~~ {\cal N}_u^2 = \sum_m\sum_{kl} |A^m_{kl}|^2 \label{norm}
\end{equation}
and $A^m_{kl}$ are the amplitudes of the process; $m$ denotes the spin quantum number of the unobserved final particle ($c$ or $d$) and $l$ and $l'$ indicate either the spin quantum number of $R_0$ or the pair ($l_a$, $l_b$) of the initial particles $a$ and $b$. 

If the number of independent amplitudes is $\bar{N}$, we parametrize them as $b_s e^{i\phi_s}$,  $s$ = 1,2, ... $\bar{N}$, with $\phi_s$ real numbers, $\phi_1$ = 0 and
\begin{eqnarray} 
b_1 &=& cos\beta_1, ~~~~~~ \ ~~~~~~ b_2 = sin\beta_1 cos\beta_2, ~~~~ ... \\
b_m &=& cos\beta_m \Pi_{l=1}^{m-1} sin\beta_l ~~~ ... ~~~ b_{\bar N} = \Pi_{l=1}^{\bar{N}-1} sin\beta_l. 
\end{eqnarray}
Then the amplitudes are characterized by $2({\bar N}-1)$ real parameters, to which one has to add those of the initial SDM. Obviously, if parity is conserved, one has to take account of the corresponding symmetry relations. 

In order to see whether this method is more convenient than the previous one, the overall number of such parameters has to be compared with the upper bound to the rank of the density matrix of $R$ is $r_i\times r_{c(d)}$, where $r_{i(c,d)}$ are respectively the ranks of the initial state and of $c$ or $d$. In sect. 6, we shall compare the two methods in some cases.    

\section{Discussion}

The number of independent parameters, that one can extract from the normalized differential decay width, is determined by the moments, {\it i. e.},
\begin{equation}
\frac{1}{\Gamma}\frac{d^2\Gamma}{d cos\theta d\phi} =
\sum_{L=0}^{J_m}\sum_{M=-L}^L \frac{2L+1}{4\pi} H(L,M) {\cal D}^{L*}_{M0}(\theta,\phi).
\end{equation}
Here the ${\cal D}$ are the Wigner rotation functions and $J_m$ is the maximum spin of $R$. Moreover, one has $H(0,0)$ = 1. The other moments can be extracted either by means of a best fit or by exploiting the orthogonality of the ${\cal D}^L_{M0}$ functions with different $L$ or $M$. 

If $R$ includes only one spin, the number of parameters of the SDM equals the number of moments, which are related to them by a determined linear system\cite{cht,chu}. Otherwise, the best we can extract are a number of matrix elements and some linear combinations of the remaining ones\cite{cht}, unless one can reduce the number of the independent parameters of the SDM, by means of some assumption. 

If parity invariance holds both in the production and in the decay of $R$, the imaginary parts of the elements $\rho_{ij}$  of the SDM turn out to be measurable only under very particular conditions, as shown in Appendix B: 

- $R$ includes more spins and

- there are at least two independent decay amplitudes, with a non-trivial relative phase, owing to $T$-odd final-state interactions\cite{ajd}.

Such conditions are realized, for example, in the low-energy scattering $p$-$\bar{p}$ $\to$ $\Lambda$-$\bar{\Lambda}$ \cite{te}, which involves a few partial waves; the corresponding decay amplitudes have different phases, owing to
the $T$-odd (spin-orbit) term of the interaction between the final hadrons. 
 
\section{Examples}

Here we apply our parametrizations to some concrete situations, either with parity conservation or with parity violation. 

\subsection{Two Spins: $1^-$ and $0^{\mp}$ - Parity Conservation}

We consider the two cases, in order to simulate a resonance-background interference in strong decays of vector mesons; these were already considered many years ago\cite{cht}, but with a different approach. One has to take account of the relation (\ref{yr}).

a) In the case that $R$ includes the states $1^-$ and $0^-$, we have $\eta'$ = $\eta$, therefore 
\begin{equation} 
\rho^{JJ'}_{mm'} = (-)^{J-J'-m+m'} \rho^{~J~J'}_{-m-m'};
\end{equation}
this entails the following parametrization: 
\begin{eqnarray} 
\rho^{11}_{11} &=& \rho^{~1~~1}_{-1-1} = 1/2 sin^2 \alpha_1, ~~~ \rho^{11}_{00} = cos^2 \alpha_1 cos^2 \alpha_2, \label{r1}
\\
\rho^{11}_{10} &=& -\rho^{~11}_{-10} = \frac{1}{\sqrt{2}}cos\alpha_1 sin\alpha_1 cos \alpha_2 cos\gamma_{10}e^{i\varphi_{10}^{11}},\label{r2}
\\
\rho^{1~1}_{1-1} &=& \rho^{~11}_{-11} = 1/2 sin^2 \alpha_1 cos\gamma_{1-1}, ~~~ \rho^{00}_{00} = cos^2 \alpha_1 sin^2 \alpha_2,
\label{r3}
\\
\rho^{10}_{10} &=& \rho^{~10}_{-10} = \frac{1}{\sqrt{2}}cos\alpha_1 
sin\alpha_2 cos\gamma^{10}_{10}e^{i\varphi^{10}_{10}}, ~~~ \rho^{10}_{00} = 0.  \label{r4}
\end{eqnarray}	 
The remaining terms can be deduced from the previous expressions by taking account of the Hermitian character of $\rho$. Therefore, we have 7 parameters in all.

b) Consider now the case when $R$ consists of the $1^-$ and $0^+$ states. Eqs. (\ref{r1}) to (\ref{r3}) hold still true,
whereas 
\begin{eqnarray} 
\rho^{10}_{10} &=& -\rho^{~1~0}_{-10} = \frac{1}{\sqrt{2}}cos\alpha_1 
sin\alpha_2 cos\gamma^{10}_{10}e^{i\varphi^{10}_{10}},\label{rr4}
\\
\rho^{10}_{00} &=& \frac{1}{\sqrt{2}}cos\alpha_1 sin\alpha_2 cos\gamma^{10}_{00}e^{i\varphi^{10}_{00}}. \ ~~~ \ ~~~ \label{rr5}
\end{eqnarray}
Therefore, 2 more parameters are needed.

\subsection{$\Lambda_b ~ \to ~ \Lambda ~ J/\psi$}

We consider the SDM of the $J/\psi$-resonance. In this case, no symmetry holds in the $\Lambda_b$ decay, but the constraint of the angular momentum conservation has to be considered. Indeed, adopting the helicity formalism, one has 
\begin{equation}
\rho_{1-1} = \rho_{-11} = 0, \label{cstr}
\end{equation}
since the third component of the angular momentum of $\Lambda_b$ along 
the $J/\psi$ momentum assumes a fixed value, either $+1/2$ or $-1/2$, but not both simultaneously. The  other elements read as
\begin{eqnarray} 
\rho_{11} &=&   cos^2 \alpha_1, ~~
\rho_{00} = sin^2 \alpha_1 cos^2 \alpha_2, ~~ \rho_{-1-1} = sin^2 \alpha_1 sin^2 \alpha_2, \label{w1}
\\
\rho_{10} &=& \rho^*_{01} = cos\alpha_1 sin\alpha_1 cos \alpha_2 \sin^2\gamma_{10}e^{i\phi_{10}}, \label{w2}
\\
\rho_{-10} &=& \rho^*_{0-1} = sin^2\alpha_1 cos \alpha_2 \sin^2\gamma_{-10}e^{i\phi_{-10}}. \label{w3}
\end{eqnarray}
We have employed 6 parameters in all, one less than required in ref. 39, which is cited in the LHCb analysis of the decay\cite{lhca}. It is worth noting that, according to the discussion of the previous section, all such parameters can be determined, in principle, by analyzing, {\it e. g.}, the decay $J/\psi$ $\to$ $\mu^+ ~ \mu^-$. 

Incidentally, the constraint (\ref{cstr}) holds true for any vector boson resonance $V$ that comes from a decay of the type 
\begin{equation} 
f_1 ~~ \to ~~ f_2 ~~ V, \label{dcy}
\end{equation}
where $f_1$ and $f_2$ have spin 1/2. A similar simplification occurs, {\it e. g.}, for the SDM of $\Delta$(1232) in the decay 
\begin{equation} 
\Delta(1620)1/2^- ~~ \to ~~ \Delta(1232)3/2^+ ~~~~ \pi,
\end{equation}
for which parity invariance must be taken into account. 

\subsection{$t \to b W \to b \tau \nu_{\tau}$}

In this decay, which is of the type (\ref{dcy}), we employ the alternative method described in Sect. 4. We consider, in this case, the SDM of the $W$-boson emitted in a given direction, adopting a frame at rest with respect to the top quark. It reads as\cite{dsa}
\begin{equation} 
\rho_{\mu\mu'}(\theta,\phi) = \frac{1}{4\pi}\sum_{\lambda} b_{\lambda\mu} \rho'_{\Lambda\Lambda'}(\theta,\phi) b^*_{\lambda\mu'}. 
\end{equation}
Here $\Lambda$ = $\mu-\lambda$ is the spin component of $t$ along the $W$ momentum and $\mu$ and $\lambda$ the helicities of, respectively, $W$ and the $b$-quark; moreover,
\begin{eqnarray} 
\rho'_{\Lambda\Lambda'}(\theta,\phi) &=& \delta_{\Lambda\Lambda'} + \sigma^i_{\Lambda\Lambda'}P'_i,
\\
{\bf P}' &\equiv& P (cos\theta, sin\theta cos\phi, sin\theta sin\phi); 
\end{eqnarray}
$P$ is the top quark polarization, $0 \leq P \leq 1$, while  $\theta$ and $\phi$ are, respectively, the polar and the azimuthal angle of the momentum of the $W$-boson. Last, $b_{\lambda\mu}$ are the reduced decay amplitudes of the $t\to W b$ decay:
\begin{equation} 
b_{\lambda\mu} = {\cal N}^{-1/2}A_{\lambda\mu}, ~~~ {\cal N}= |A_{+,1}|^2
+|A_{+,0}|^2+|A_{-,0}|^2+|A_{-,-1}|^2.  
\end{equation}
This suggests the parametrization 
\begin{eqnarray} 
b_{+,1} &=& |cos\beta_1|, \ ~~~ \ ~~~ \ ~~~ \ \ ~~~ \ ~~~ \ ~~~ \
\\
b_{+,0} &=& |sin\beta_1 cos\beta_2| e^{i\varphi_{10}} \ ~~~ \ ~~~ \ ~ \ ~~~ \
\\
b_{-,0} &=& |sin\beta_1 sin\beta_2 cos\beta_3|e^{i\varphi_{-10}}, \ ~~~ \ ~~~ 
\\
b_{-,-1} &=& |sin\beta_1 sin\beta_2 sin\beta_3|. \ ~~~ \ ~~~ \ ~~~ \
\end{eqnarray}
Moreover, we set
\begin{equation} 
P = cos^2\beta_0.  
\end{equation}
We have assumed the phases of the amplitudes $b_{\pm,\pm 1}$ to be zero, because, in our case, they can interfere only with $b_{\pm,0}$
respectively.

The parametrization of the $W$-boson SDM results in  ${\rho}_{1-1}$ = 0 and
\begin{eqnarray} 
\bar{\rho}_{11} &=& cos^2\beta_1(1+cos^2\beta_0 cos\theta), ~~~ \bar{\rho}_{00} = sin^2\beta_1 (cos^2\beta_2 + sin^2\beta_2 cos^2\beta_3), 
\\
\bar{\rho}_{-1-1} &=& sin^2\beta_1 sin^2\beta_2 sin^2\beta_3
(1-cos^2\beta_0cos\theta), \ ~~~ \ ~~~ \ ~~~ \ ~~~ \ ~~~ \
\\
\bar{\rho}_{10} &=&|cos\beta_1 sin\beta_1 cos\beta_2| 
(1+ cos^2\beta_0 sin\theta e^{-i\phi}) e^{i\varphi_{10}}, 
\ ~~~ \ ~~~ \ ~~~ ~~~ \
\\	
\bar{\rho}_{-10} &=& |sin\beta_1 sin\beta_2|(|cos\beta_3|+|sin\beta_3|)
(1+ cos^2\beta_0 sin\theta e^{i\phi}) e^{i\varphi_{-10}}, \ ~~~ \ ~~~ 
\ ~~~ ~~~ \ 
\end{eqnarray}
having set $\bar{\rho}$ = $4\pi\rho$. As before, the remaining SDM elements are deduced from the hermiticity condition.

It is worth noting that this procedure - whose parameters, again, do not need any bound - could also be applied to the vector boson $V$ in a decay of the type (\ref{dcy}); it has the advantage of relating the polarization of the $W$ to that of the parent state.

\subsection{$ \Lambda_b \to \Lambda_c (W^*,H^*) \to \Lambda_c \tau \nu_{\tau}$}

In this case, the independent (reduced) decay amplitudes are  8:
\begin{eqnarray} 
b^1_{+,1} &=& |cos\beta_1|, ~~~ b^1_{+,0} = |sin\beta_1 cos\beta_2| e^{i\varphi^1_{+,0}}, ~~~ b^1_{-,0} = |sin\beta_1 sin\beta_2 cos\beta_3| e^{i\varphi^1_{-,0}},
\\
b^1_{-,-1} &=& |sin\beta_1 sin\beta_2 sin\beta_3 cos\beta_4|, ~~~
b^t_{+,0} = \prod_{k=1}^4 |sin\beta_k cos\beta_5| e^{i\varphi^t_{+,0}}, 
\\
b^t_{-,0} &=& \prod_{k=1}^5 |sin\beta_k cos\beta_6| e^{i\varphi^t_{-,0}}, ~~~ b^0_{+,0} = \prod_{k=1}^6 |sin\beta_k cos\beta_7| e^{i\varphi^0_{+,0}},
\\
b^0_{-,0} &=& \prod_{k=1}^7 |sin\beta_k| e^{i\varphi^0_{-,0}}. 
\end{eqnarray}
Note that the virtual character of the $W$ implies 4 vector amplitudes, $b^1_{\pm,0}$, $b^1_{+,1}$ and $b^1_{-,-1}$, and 2 scalar ones\cite{dsa}, $b^t_{\pm,0}$; the possible new physics is described by the 2 remaining amplitudes, $b^0_{\pm,0}$. Moreover, analogously to the previous case, we have assumed equal to 0 the phases of $b^1_{\pm,\pm1}$.

Then, following the same procedure and using the same notations as in the previous subsection, we find that the matrix elements of $\bar{\rho}$ read as    
\begin{eqnarray} 
\bar{\rho}^{1~1}_{1-1} &=& 0, ~~~ \bar{\rho}^{~1~1}_{\pm1\pm1} = \xi_{\pm1}(1\pm P_z), 
\\
\bar{\rho}^{JJ'}_{0~0} &=& \eta^{JJ'}_{00}+ \Delta\eta^{JJ'}_{0~0} P_z ~~~ \bar{\rho}^{~1J}_{\pm1 0} = \eta^{~1~J}_{\pm1 ~0} P_{\mp}.  
\end{eqnarray}
Here $J$ and $J'$ run over 1, $t$, 0; moreover,
\begin{equation}
P_z = cos^2\beta_0 cos\theta, ~~~ P_{\pm} = cos^2\beta_0 sin\theta e^{\pm i\phi};
\end{equation}
last,
\begin{eqnarray} 
\xi_{\pm1} &=& |b^1_{\pm,\pm1}|^2, ~~~ \eta^{JJ'}_{0~0} = b^J_{+,0} b^{J'*}_{+,0}+ b^J_{-,0} b^{J'*}_{-,0},
\\ 
\Delta\eta^{JJ'}_{00} &=& b^J_{-,0} b^{J'*}_{-,0}- b^J_{+,0} b^{J'*}_{+,0}, ~~~ \eta^{~1J}_{\pm1 0} = b^1_{\pm,\pm1} b^{J*}_{\pm,0}.
\end{eqnarray}
In this case, we have used 14 parameters in all, whereas the general treatment would require 24 parameters. But according to the considerations of Sect. 5 the number of independent parameters that can be inferred from the differential decay width is 8; therefore, the best fit to the data, if performed by using the above parametrization,
would present some  ambiguities. However, in this case, some parameters can be fixed by inserting the standard model predictions\cite{ab1,ab2,fi2,fi3}. In particular, in a previous paper\cite{dsa}, we showed the relationship between the tensor, that is usually employed to describe a semi-leptonic decay, and the non-covariant SDM. If at least one of the amplitudes $b^0_{\pm,0}$ is non-zero, with a non-trivial phase, it causes a T-odd component for the $\Lambda_c$ polarization\cite{ad17}.

\section{Conclusions}

We have proposed two methods for parametrizing the SDM of an unstable state - consisting of one or more spins - that is produced in various reactions. Both methods satisfy automatically all of the numerous non-negativity conditions\cite{da,min} and are adaptable to the constraints imposed by parity\cite{cht} and angular momentum conservation.
The first method is based on a theorem that we have proved and it may be applied in a simple and flexible way. Moreover, we show how to improve the fit to the data when the rank of the SDM is less than its order; we do this, either by implementing suggestions by other authors\cite{min,bls}, or with the help of a second theorem, which we have proved as well.

The second method is a variant of previous parametrizations\cite{pe,bls} and is particularly suitable under especial conditions, {\it e. g.}, when the structure to be analyzed derives from some decay.

We have discussed about the possibility of inferring the SDM from the differential decay width. In particular, we have
examined the case where a given intermediate state is generated and decays according to strong or electromagnetic interactions, showing that, under very particular circumstances, the imaginary part of the SDM can be measured. 

Last, we have illustrated some applications; two of them, which concern the decays $t ~ \to b ~ W$\cite{ab1,ab2} and $\Lambda_b ~ \to ~ \Lambda_c ~ l^+ ~ \nu_l$[44-46,22], are very interesting from the viewpoint of the search for new physics beyond the Standard Model. 

As a conclusion, we observe that our suggestions appear efficient for situations where higher spins or newly discovered structures\cite{lhcb} are involved. 

\vskip 0.25in

\centerline{\bf Acknowledgments}
The authors are thankful to their friend Flavio Fontanelli for useful and stimulating discussions.
 
\vskip 0.25in 
\setcounter{equation}{0}
 \renewcommand\theequation{A. \arabic{equation}}
\appendix{\large \bf Appendix A}
\vskip 0.30cm 
We find an explicit solution to the $N$ complex vectors $|W_{i}\rangle$ which appear in Eq. (\ref{scpr1}), {\it i. e.},
\begin{equation} 
\rho_{ij} = \langle W_{i}|W_{j}\rangle. \label{meq}
\end{equation}
This is possible owing to the Schwarz inequality, that we have assumed.
The data of the system (\ref{meq}) consist of the elements of the hermitian matrix $\rho$, defined with respect to an 
orthonormal basis $|k\rangle$ and such that 
\begin{equation} 
\rho_{ii} \geq 0 \ ~~~ \ \mathrm{and} \ ~~~ \ |\rho_{ij}|^2 \leq \rho_{ii}\rho_{jj}.
\end{equation}

We assume expansions of the type
\begin{equation} 
|W_i\rangle = \sum_{k=1}^i \alpha^k_i |k\rangle  \label{exps}
\end{equation}
for any vector $|W_{i}\rangle$, $i=1,2,..N$; furthermore, we establish all of the $\alpha^i_i$ to be real and non-negative. 
Then, limiting ourselves to $i \geq j$, Eq. (\ref{meq}) yields
\begin{equation} 
\rho_{ij} = \sum_{k=1}^i {\alpha^k_i}^* \alpha^k_j. \label{sol}
\end{equation}
We show that this equation uniquely fixes all of the coefficients $\alpha^k_i$ of the expansions (\ref{exps}).

1) Assume, at first, that the rank of $\rho$ is equal to $N$, which implies that all principal minors
of the matrix are positive, in particular, $\rho_{ii}$ $>$ 0. Then we prove our statement by induction.

a) For $i$ =1, Eq. (\ref{sol}) reads as 
\begin{equation} 
\alpha^1_1 = \sqrt{\rho_{11}}, \label{sol1}
\end{equation}
which defines $|W_1\rangle$ through Eq. (\ref{exps}).

b) Suppose Eqs. (\ref{sol}) to be solvable with respect to $\alpha^k_j$ for all $k\leq j$ and all $j\leq i<N$.
This amounts to asserting that all vectors   
\begin{equation} 
|W_j\rangle, \ ~~~ \ j \leq i, \label{bas}
\end{equation}
have been determined. Now we prove that the vector
\begin{equation} 
|W_{i+1}\rangle = \sum_{k=1}^{i+1} \alpha^k_{i+1} |k\rangle,
\end{equation}
can  be uniquely deduced from the system 
\begin{equation} 
\rho_{i+1j} = \sum_{k=1}^j {\alpha^{k*}_{i+1}} \alpha^k_j, \ ~~~ \ j=1,2,..i+1, \label{solj}
\end{equation}
a complex system of $i+1$ equations, with $i$ complex unknowns, $\alpha^k_{i+1}$, $1 \leq k \leq i$, and a real one,
$\alpha^{i+1}_{i+1}$.

To this end, preliminarily, we show that the $i\times i$ ('triangular') matrix $A$, such that 
\begin{equation} 
A_{jk} = \alpha^k_j, \ ~~~ \ \mathrm{with} \ ~~~ \ k \leq j \leq i,  \label{syst}
\end{equation}
is non-singular. 
Consider the $i\times i$ submatrix $\tilde{\rho}$, whose matrix elements $\tilde{\rho}_{lm}$ coincide with those 
of $\rho$ for $l,m$ $\leq$ $i$. Eq. (\ref{meq}) implies
\begin{equation} 
\tilde{\rho}_{lm} = \langle W_l|W_m\rangle. \label{meqt}
\end{equation}
The assumption of non-singularity of $\rho$ implies the same for $\tilde{\rho}$, therefore the vectors (\ref{bas}) 
constitute a basis for $\tilde{\rho}$ and are linearly independent. This in turn entails the non-singularity of
$A$\footnote{As a byproduct, it is worth noting that, in this case, all $\alpha_i^i$ are strictly positive.}. 

But the system (\ref{solj}) can be split into a linear subsystem with $i$ equations and a non-linear equation: 
\begin{equation} 
\rho_{i+1j} = \sum_{k=1}^j A_{jk} {\alpha^{k*}_{i+1}}, \ ~~~ \ j=1,2,..i, \ ~~~ \ 
\rho_{i+1i+1} = \sum_{k=1}^{i+1}|\alpha^k_{i+1}|^2.
\end{equation}
The non-singular character of $A$ allows to solve the linear subsystem with respect to $\alpha^{k}_{i+1}$, $k$ = $1,2,..i$. The solution can be inserted into the non-linear equation, which can be solved with respect to $\alpha^{i+1}_{i+1}$. This completes the proof in the case of non-singular $\rho$.

2) If the rank of $\rho$, say $r$, is less than $N$, we perform the transformation (\ref{tdm}), {\it i. e.}   
\begin{equation} 
\rho' = U \rho U^{\dagger}.  \label{unit}
\end{equation}
The result is 
\begin{equation} 
{\rho'}_{lm} = \rho_{lm} = \eta_{lm} \ ~~~ \ \mathrm{for} \ ~~~ \ 1 ~ \leq ~ l,m ~ \leq r  \ ~~~ \ \mathrm{and} \ ~~~ \ 0 \ ~~~ \ \mathrm{otherwise};
\end{equation}
therefore $\eta$ is an $r\times r$ Hermitian, non-singular matrix, to which we may apply the method described above. We find a set of linearly independent complex vectors $|w_l\rangle$, $l$ = 1, 2,...$r$. Defining 
\begin{equation} 
|W'_i\rangle = |w_i\rangle  ~~~ \mathrm{for} ~~~ i \leq r ~~~  \mathrm{and} ~~~ 0 
 ~~~  \mathrm{for} ~~~ r < i \leq N,
\end{equation}
the transformed vectors
\begin{equation} 
|W_i\rangle = U_{ij}|W'_j\rangle 
\end{equation}
give the solution to the system (\ref{meq}). This completes the proof.

\vskip 0.25in 
\setcounter{equation}{0}
 \renewcommand\theequation{B. \arabic{equation}}
\appendix{\large \bf Appendix B}
\vskip 0.30cm

Here we discuss about the measurability of the imaginary parts of the  elements of the spin density matrix (SDM) of an unstable state $R$ that is produced and decays according to parity conserving interactions. We assume this state to include more spins and to have a two-body decay:
\begin{equation} 
R ~~ \to ~~ a ~~ b.
\end{equation}
Adopting the helicity formalism, the normalized differential decay width reads as
\begin{equation} 
\frac{1}{\Gamma}\frac{d^2\Gamma}{d cos\theta d\phi} = \sum_{J J'} 
\sum_{m m'} \sum_{\lambda_a\lambda_b} C(J,J') \rho_{m m'}^{J J'}{\cal D}^{J*}_{m\lambda}(\phi,\theta,0)
{\cal D}^{J'}_{m'\lambda}(\phi,\theta,0)f^J_{\lambda_a\lambda_b} f^{J'*}_{\lambda_a\lambda_b}. \label{ad}
\end{equation}	
Here $C(J,J')$ = $1/4\pi [(2J+1)(2J'+1)]^{1/2}$, $\rho_{m m'}^{J J'}$ = $\langle J,m|\rho|J'm'\rangle$ is an element of the SDM, ${\cal D}$ is the Wigner rotation function and 
\begin{equation} 
f^J_{\lambda_a\lambda_b} = \frac{1}{N_f} F^J_{\lambda_a\lambda_b},
~~~~~~ N_f^2 = \sum_{J,\lambda_a,\lambda_b} | F^J_{\lambda_a\lambda_b}|^2,
\end{equation}
the reduced decay amplitudes, with $\lambda$ = $\lambda_a-\lambda_b$. Parity conservation implies
\begin{eqnarray} 
\rho_{m m'}^{J J'} &=& \eta \eta' e^{-i\pi\Delta}\rho_{-m -m'}^{~J ~~J'}, ~~~~~~ \Delta=J-J'-m+m', \label{dmsy}
\\
f^J_{-\lambda_a-\lambda_b} f^{J'*}_{-\lambda_a-\lambda_b} &=& \eta \eta' e^{-i\pi(J-J')} \label{dcsy}
f^J_{\lambda_a\lambda_b} f^{J'*}_{\lambda_a\lambda_b},
\end{eqnarray}
as follows from Eqs. (\ref{stpp}) to (\ref{yr}) in the text. 
Moreover, 
\begin{equation}
{\cal D}^J_{m\lambda}(\phi,\theta,0)= e^{-im \phi}d^J_{m\lambda}(\theta), ~~~~~~ d^J_{-m-\lambda}(\theta) = (-)^{m-\lambda} d^J_{m\lambda}(\theta).
\end{equation}
Therefore we may re-write Eq. (\ref{ad}) as
\begin{eqnarray} 
\frac{1}{\Gamma}\frac{d^2\Gamma}{d cos\theta d\phi} = \sum_{J J'} 
{\sum_{m m'}}' {\sum_{\lambda_a\lambda_b}}' C(J,J') [e^{-i(m-m')\phi} +(-)^{\epsilon} e^{i(m-m')\phi}] \nonumber \\
\times \rho_{m m'}^{J J'} d^J_{m\lambda}(\theta)d^{J'}_{m'\lambda}(\theta)f^J_{\lambda_a\lambda_b} f^{J'}*_{\lambda_a\lambda_b}, \label{adp}
\end{eqnarray}
where $\epsilon = \Delta+J-J'+m+m'-2\lambda$ = $2(J-J'+m'-\lambda)$ and the primes in the sums indicate that they are limted to non-negative values of the indices. But 
$\epsilon$ is an even number, moreover only $\Im \rho_{m m'}^{J J'}$ and $\Im (f^J_{\lambda_a\lambda_b} f^{J'*}_{\lambda_a\lambda_b})$ are odd under the simultaneous exchange $(J,m)$ $\leftrightarrow$ $(J',m')$, the other terms of Eq. (\ref{adp}) being even.
Therefore we have 
\begin{equation} 
\frac{1}{\Gamma}\frac{d^2\Gamma}{d cos\theta d\phi} = \sum_{J J'} 
{\sum_{m m'}}' {\sum_{\lambda_a\lambda_b}}' C(J,J') cos[(m-m')\phi] d^J_{m\lambda}(\theta)d^{J'}_{m'\lambda}(\theta) 
{\cal F}^{J J'}_{m m',\lambda_a\lambda_b}, \label{adf}
\end{equation}
with
\begin{equation}
{\cal F}^{J J'}_{m m',\lambda_a\lambda_b} = \Re\rho_{m m'}^{J J'} \Re(f^J_{\lambda_a\lambda_b} f^{J'*}_{\lambda_a\lambda_b})-
\Im\rho_{m m'}^{J J'} \Im(f^J_{\lambda_a\lambda_b} f^{J'*}_{\lambda_a\lambda_b}).
\end{equation}
Therefore, under the conditions that we have assumed, the imaginary part of the SDM can be measured only if more spins are involved and at least one of the relative phases between the decay amplitudes is non-trivial.

\end{document}